\documentclass[10pt,prl,superscriptaddress,preprintnumbers,balancelastpage,twocolumn]{revtex4-2}

\usepackage{graphicx}
\usepackage{epstopdf}
\usepackage{bm}
\usepackage[utf8]{inputenc}
\PassOptionsToPackage{hyphens}{url}
\PassOptionsToPackage{breaklinks=true}{hyperref}
\usepackage[hidelinks,colorlinks]{hyperref}
\usepackage{soul}
\usepackage{amsmath}
\usepackage{amssymb}
\usepackage{mathtools}
\usepackage{xfrac}
\usepackage{rotating}
\usepackage{enumitem}
\usepackage{scrextend}
\usepackage{slashed}
\usepackage{braket}
\usepackage{dcolumn}
\usepackage{caption}
\usepackage{subcaption}
\usepackage[dvipsnames]{xcolor}
\usepackage{tikz}
\usetikzlibrary{calc}
\usetikzlibrary{arrows.meta, positioning}

\usepackage{url} 
\hypersetup{
    colorlinks=true,
    linkcolor=blue,
    citecolor=red,
    urlcolor=olive
}

\def\Hhard{\mathcal{H}} 

\def\MM{\mathcal{M}}

\newcommand{\bea}{\begin{eqnarray}}
\newcommand{\eea}{\end{eqnarray}}
\newcommand{\beqa}{\begin{eqnarray}}
\newcommand{\eeqa}{\end{eqnarray}}
\newcommand{\be}{\begin{equation}}
\newcommand{\ee}{\end{equation}}
\newcommand{\beq}{\begin{equation}}
\newcommand{\eeq}{\end{equation}}

\begin{document}


\title{
Infrared singularities of multileg amplitudes with a massive particle at three loops
}

\author{Einan Gardi}
\email{Einan.Gardi@ed.ac.uk}
\author{Zehao Zhu}%
 \email{Z.Zhu-37@sms.ed.ac.uk}
\affiliation{%
 Higgs Centre for Theoretical Physics, School of Physics and Astronomy,\\ The University of Edinburgh, Edinburgh EH9 3FD, Scotland, UK
}%

\date{\today}

\begin{abstract}
We determine the complete three-loop QCD soft anomalous dimension for multileg amplitudes involving a single massive coloured particle and any number of massless ones. This is achieved by applying a novel strategy based on 
a lightcone expansion of  
correlators of semi-infinite Wilson lines  using the method of regions. The resulting region integrals depend exclusively on rescaling-invariant ratios that remain finite in the limit. We evaluate these integrals using differential equation techniques. The result is written in terms of uniform weight five generalised polylogarithms of a twelve letter alphabet in three variables, and is compatible with the massless limit as well as with two- and three-particle collinear factorization. 
\end{abstract}

\maketitle

Infrared singularities are an important feature of on-shell gauge-theory amplitudes.
These singularities admit a highly-constrained, universal structure~\cite{Polyakov:1980ca,Arefeva:1980zd,Dotsenko:1979wb,Brandt:1981kf,Sen:1982bt,Korchemsky:1985xj,Ivanov:1985np,Korchemsky:1987wg,Korchemsky:1988hd,Korchemsky:1991zp,Korchemskaya:1994qp,Catani:1998bh,Kidonakis:1998nf,Kidonakis:1997gm,Sterman:2002qn,Gardi:2005yi,Aybat:2006mz,Dixon:2008gr,Gardi:2009qi,Gardi:2009zv,Becher:2009cu,Becher:2009qa,Ma:2019hjq,Feige:2014wja,Agarwal:2021ais,Frenkel:1984pz,Gatheral:1983cz,Sterman:1981jc,Mitov:2010rp,Gardi:2010rn,Gardi:2011wa,Gardi:2011yz,Gardi:2013ita,Agarwal:2020nyc,Almelid:2015jia,Almelid:2016lrq,Almelid:2017qju,Becher:2019avh,Falcioni:2021buo,Maher:2023jqy}. Their origin in soft and collinear loop momentum regions guarantees their all-order factorization from the finite hard amplitude. Using renomalization-group equations, factorization can be used to prove that the singularities of any amplitude exponentiate in terms of the so-called soft anomalous dimensions. This universal quantity is therefore of unique importance in the study of amplitudes~\cite{Agarwal:2021ais}.   

In QCD,  infrared singularities present one of the key challenges for collider physics. Their cancellation in observables involves a sum over states of different multiplicity, requiring sophisticated subtraction techniques~\cite{DelDuca:2024ovc,Alioli:2025hpa,Currie:2013vh,Chargeishvili:2021pxd,Magnea:2024jqg,Magnea:2024jqg,Devoto:2023rpv,Devoto:2025kin}. The singularities also give rise to enhanced logarithmic corrections, which often need to be resummed to achieve precise predictions. 
In any of these applications, the soft anomalous dimension is essential.

The soft anomalous dimension for multileg massless scattering is currently known to three loops~\cite{Almelid:2015jia}. However, in the case of massive particles, the state of the art is still two loops~\cite{Ferroglia:2009ep,Ferroglia:2009ii}~\footnote{The cusp anomalous dimension, controlling dipole corrections is known in full to three loops ~\cite{Grozin:2014hna,Grozin:2015kna}.}.
Important collider physics applications, such as resummation of top production cross sections~\cite{Czakon:2013hxa,Angeles-Martinez:2018mqh,Shao:2025qgv,Buonocore:2025ysd,Liu:2024hfa,Kidonakis:2025eia,Guzzi:2024lol,Kidonakis:2024lht,Kidonakis:2023juy}, require knowledge of the three-loop soft anomalous dimension for processes involving heavy quarks alongside light quarks and gluons. Recently, first results for such corrections were obtained~\cite{Liu:2022elt}, namely those associated with the interaction between one massive particle and two massless ones. In present paper we compute the remaining contributions, and provide the complete three-loop result for the soft anomalous dimension in 
amplitudes involving one 
massive particle and any number of massless ones. 

The soft anomalous dimension is remarkably simple in comparison with finite corrections to multileg amplitudes. In particular, its multileg components are spin-independent and feature rescaling symmetry with respect to the external momenta, as expected from its relation with Wilson-line correlators.
This symmetry significantly reduces the number of independent kinematic variables on which the soft anomalous dimension depends, especially so in  massless scattering~\cite{Becher:2009cu,Gardi:2009qi,Becher:2009qa,Dixon:2009gx,Gardi:2009zv}. However, it has been difficult to make use of this simplification when computing the soft anomalous dimension, because multiplicative renormalizabilty~\cite{Polyakov:1980ca,Arefeva:1980zd,Dotsenko:1979wb,Brandt:1981kf} of Wilson-line correlators only holds when the Wilson lines are non-lightlike. 
Our novel strategy for computing the soft anomalous dimension~\cite{Gardi:2025ule} directly addressed this issue. We use the method of regions (MoR)~\cite{Beneke:1997zp,Smirnov:2002pj,Pak:2010pt,Jantzen:2011nz,Jantzen:2012mw,Semenova:2018cwy,Ananthanarayan:2018tog,Heinrich:2023til,Borowka:2017idc,Gardi:2022khw,Ma:2023hrt,Chen:2024xwt} to perform an asymptotic  lightcone expansion of  correlators of timelike Wilson lines. In this way, the integrals we compute involve only the kinematic variables that remain finite in the limit of interest.

\section{One-mass soft anomalous dimension}
Consider a renormalized amplitude $\MM$ with one massive particle $p_Q$ and any number of massless particles~$\{p_i\}$~\footnote{We use upper-case letters, such as $Q$, to label massive particles and timelike Wilson lines, and lower case, e.g.~$i,j,k$, to label massless particles and lightlike Wilson lines. }. Infrared singularities can be factorized as $\MM\, = Z \Hhard$ where the singularities are given by
\begin{align}
 \label{IRfacteqZ}
 \begin{split}
Z \, =
\text{P}{\displaystyle \exp}
\left\{{-\!\displaystyle\int_{0}^{\mu}\!\frac{d\tau}{\tau}{\mathbf \Gamma}\left(\tau,\alpha_s(\tau^2)\right)}\right\}\,
\end{split}
\end{align}
and ${\cal H}$ is the finite hard amplitude. The one-mass soft anomalous dimension is 
\begin{eqnarray}
    \label{cstructure}
{\mathbf{\Gamma}}
&=&\sum_{i<j}\mathbf{T}_{i}\cdot\mathbf{T}_{j}\gamma_{\text{cusp}}(\alpha_s)\log\left(-\frac{\tau^2}{2p_i\cdot p_j}\right) \,+\,\sum_{i}\gamma_i(\alpha_s)
\nonumber \\&+ &
\sum_{i}\mathbf{T}_{i}\cdot\mathbf{T}_{Q}\gamma_{\text{cusp}}(\alpha_s)\log\left(-\frac{\tau\sqrt{p_Q^2}}{2p_i\cdot p_Q}\right)\, +\, \Omega_Q(\alpha_s)
   \nonumber \\&+&
    {\color{blue}
\sum_{i<j<k<q}\sum_{(u,v;w)\in P}
\mathbf{T}_{uv;w q} {\cal F}_{0,4}(\rho_{uwv q},\rho_{vwu q })
}
\nonumber \\ &+ &\,\,
{\color{blue}
{\cal F}_{0,3}\, \sum_{i}\sum_{j<k;\,j,k\neq i}\mathbf{T}_{ij;ik}
}
\nonumber \\&+&
{\color{violet}\sum_{i<j<k}\sum_{(u,v;w)\in P}
\mathbf{T}_{uv;w Q} {\cal F}_{1,3}(r_{uw Q},r_{vw Q},r_{uvQ})}
\nonumber \\
&+
&\,\, 
{\color{violet}
\sum_{i<j}\mathbf{T}_{iQ;jQ}{\cal F}_{1,2}(r_{ijQ})}
\,\, +\,\, {\cal O}(\alpha_s^4)\,,
\end{eqnarray}
where~$\mathbf{T}_i$ is the colour generator associated with the parton $i$~\cite{Catani:1996jh}. The dipole terms entering the first two lines in Eq.~\eqref{cstructure} start at one loop, while the quadrupole ones in the subsequent lines at three loops,
\begin{align}
\begin{split}
    {\cal F}_{h,l}=\,\,&\sum_{n=3}^{\infty}\left(\frac{\alpha_s}{4\pi}\right)^n{\cal F}_{h,l}^{(n)}\,.
    \end{split}
\end{align}
Here the subscripts $h$ and $l$ count the number of heavy and light particles, respectively. 
In Eq.~(\ref{cstructure}) we suppress any contribution associated with two or more heavy particles, and any new structures appearing from four loops~\cite{Becher:2019avh,Falcioni:2021buo,Maher:2023jqy}. 
The lighlike $\gamma_{\text{cusp}}$ is known to four loops~\cite{Moch:2004pa,Henn:2016men,Davies:2016jie,Henn:2016wlm,Lee:2017mip,Moch:2017uml,Grozin:2018vdn,Moch:2018wjh,Lee:2019zop,Henn:2019rmi,vonManteuffel:2019wbj,Henn:2019swt,vonManteuffel:2020vjv,Agarwal:2021zft}, while 
the colour-singlet terms, $\gamma_i$ and $\Omega_{Q}$ are known to four loops~\cite{vonManteuffel:2020vjv,Agarwal:2021zft,Moch:2005id,Moch:2005tm,Baikov:2009bg,Becher:2006mr} and three loops~\cite{Grozin:2014hna,Grozin:2015kna,Bruser:2019yjk}, respectively~\footnote{We summarise the results through three loops in the supplemental material.}. 

\begin{figure*}
\begin{subfigure}{0.18\linewidth}
    \centering \includegraphics[width=1\linewidth]{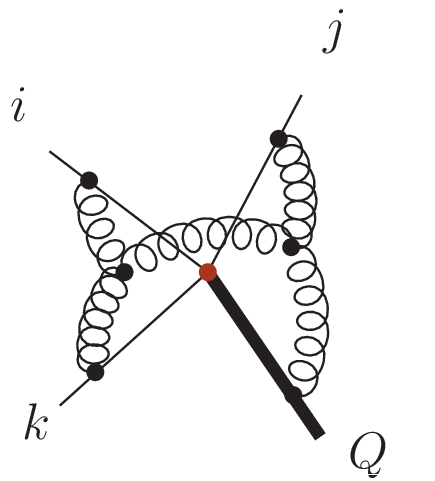}
    \label{fig:11113g3g}
    \end{subfigure}
    \hfill
    \begin{subfigure}{0.18\linewidth}
    \centering \includegraphics[width=1\linewidth]{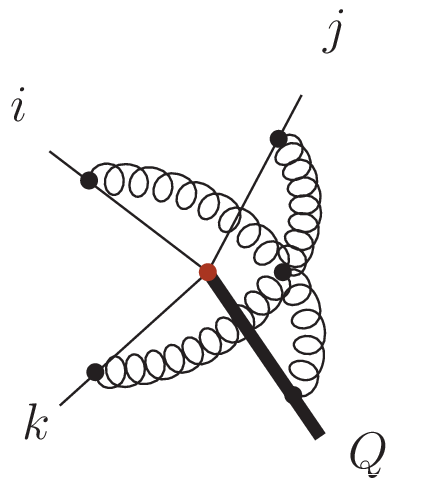}
    \label{fig:11114g}
    \end{subfigure}
    \hfill
     \begin{subfigure}{0.18\linewidth}
    \centering \includegraphics[width=1\linewidth]{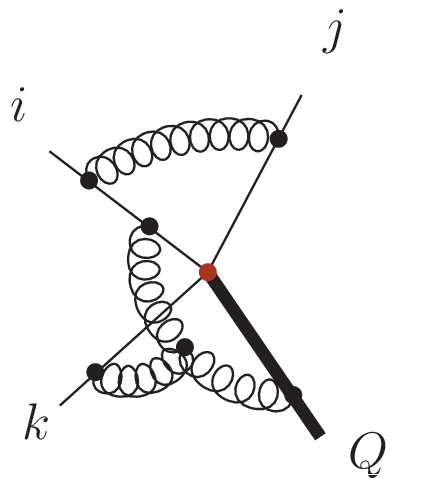}
    \label{fig:1112}
    \end{subfigure}
    \hfill
     \begin{subfigure}{0.18\linewidth}
    \centering \includegraphics[width=1\linewidth]{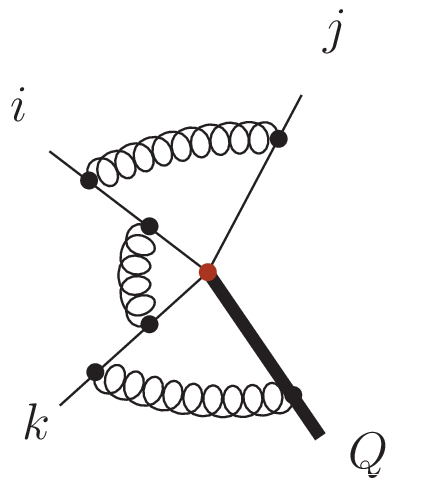}
    \label{fig:1122}
    \end{subfigure}
    \hfill
     \begin{subfigure}{0.18\linewidth}
    \centering \includegraphics[width=1\linewidth]{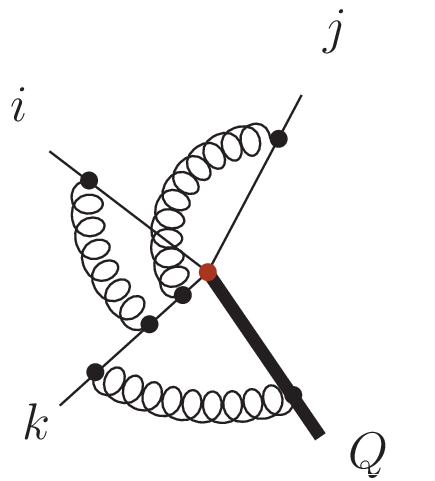}
    \label{fig:1113}
    \end{subfigure}
     \caption{Representative three-loop diagrams contributing to the quadrupole structure. The thick line corresponds to the timelike Wilson line with velocity $\beta_Q$, while the thin lines to Wilson lines with   nearly lightlike velocities. The two leftmost diagrams depict connected webs, $W_{1111}$, the middle one is a representative diagram of the $W_{1112}$ web and the two rightmost ones belong to  multiple gluon-exchange webs,~$W_{1122}$ and  $W_{1113}$, respectively.}
\label{fig:threeloop4linewebs}
\end{figure*}

In  Eq.~(\ref{cstructure})  
$(u,v;w)\in P\equiv \{(i,k;j),(j,k;i),(i,j;k)\}$ defines the three quadrupole channels, having colour structures 
\begin{equation}
\label{Tuvwq}
\mathbf{T}_{uv;w q}\equiv
f^{abe}f^{cde}
\{\mathbf{T}_{u}^a,
\mathbf{T}_{v}^b,
\mathbf{T}_{w}^c,
\mathbf{T}_{q}^d
\}_{+}
\end{equation}
where the curly brackets are defined,  as in~Refs.~\cite{Becher:2019avh,Falcioni:2021buo,Duhr:2025cye}, as a symmetric sum over all permutations of the $n$ generators it contains, divided by~$n!$~\footnote{Since generators associated with different partons trivially commute, this average over permutations only matters when two or more of the generators are associated with the same parton, as in the fourth or sixth lines of Eq.~(\ref{cstructure})}. Beyond Bose symmetry, the three channels in $P$ are related by the Jacobi identity  
 \hbox{$  f^{abe}f^{cde}+f^{ade}f^{bce}=f^{ace}f^{bde}$}.

A salient feature of the quadrupole corrections is that they depend on the momenta only via rescaling-invariant ratios~\cite{Becher:2009cu,Gardi:2009qi,Becher:2009qa,Dixon:2009gx,Gardi:2009zv,Dixon:2009ur,Almelid:2015jia} of the 4-momenta $p_u$, or equivalently the 4-velocities $\beta_u$. 
In the massless case, these are 
\begin{align}
\label{rhoDef}    \rho_{ijkq}\equiv\,\,&\,\frac{p_i\cdot p_jp_k\cdot p_q}{p_i\cdot p_kp_j\cdot p_q}=\frac{\beta_i\cdot \beta_j\beta_k\cdot \beta_q}{\beta_i\cdot \beta_k\beta_j\cdot \beta_q}\,,
\end{align}
where ${\cal F}_{0,4}$ involves just two independent cross ratios. In contrast, in the one-mass case, there are three independent rescaling-invariant ratios of the form~\cite{Liu:2022elt}
\begin{align}
\label{rDef}  r_{ijQ}\equiv \,\frac{p_Q^2p_i\cdot p_j}{2p_i\cdot p_Qp_j\cdot p_Q}
=\frac{\beta_Q^2\beta_i\cdot\beta_j}{2\beta_i\cdot\beta_Q\beta_j\cdot\beta_Q}\,.
\end{align}

The massless quadrupole terms, ${\cal F}_{0,3}$ and ${\cal F}_{0,4}$, were computed at three loops~\cite{Almelid:2015jia,Almelid:2016lrq}:
\begin{eqnarray}
 \label{Fmassless}
{\cal F}_{0,3}^{(3)}&=&\,\,32\left[2\zeta(2)\zeta(3)+\zeta(5)\right]\equiv 32C, \\ \nonumber
  {\cal F}_{0,4}^{(3)}(\rho_{ijkq},\rho_{kjiq})&=&\,\,16\left[F_{0,4}(1-\bar{z},1-z)-F_{0,4}(z,\bar{z})\right],
  \\
&&\hspace*{-80pt}
F_{0,4}(z,\bar{z})\equiv\,\,\mathcal{L}_{10101}(z)+\,2\zeta(2)\left[\mathcal{L}_{100}(z)+\mathcal{L}_{001}(z)\right],\nonumber 
\end{eqnarray}
where the variables $z$ and $\bar{z}$ are defined by
\begin{align}
\label{zzbVar}
\begin{split}
\rho_{ijkq}\equiv\,\,&z\bar{z},
\qquad
\rho_{kjiq}\equiv(1-z)(1-\bar{z})\,,
\end{split}
\end{align} 
and $\mathcal{L}(z)$ are single-valued~\cite{Brown:2004ugm,Dixon:2012yy} harmonic polylogarithms. They are single valued when $z=(\bar{z})^*$. The expansion of $F_{0,4}(z,\bar{z})$ in ordinary (multi-valued) Generalised Polylogarithms (GPL) of $z$ and $\bar{z}$ has been provided in Ref.~\cite{Almelid:2015jia}.
The one-mass three-line quadrupole, which depends on a single ratio $r_{ijQ}$, was obtained in Ref.~\cite{Liu:2022elt}, ${\cal F}_{1,2}^{(3)}(r_{ijQ})= 2 F_{h2}(r_{ijQ})$.
Thus the only unknown ingredient in Eq.~(\ref{cstructure}) prior to this work is ${\cal F}_{1,3}^{(3)}$ of the penultimate line, which describes long-distance interaction between the heavy particle and three massless ones. 
In this paper we compute ${\cal F}_{1,3}^{(3)}$ using a novel method based on the lightcone expansion of Wilson-line correlators; representative diagrams are shown in Fig.~\ref{fig:threeloop4linewebs}~\footnote{Following previous work on diagrammatic exponentiation in QCD~\cite{
Frenkel:1984pz,Gatheral:1983cz,Sterman:1981jc,Mitov:2010rp,Gardi:2010rn,Gardi:2011wa,Gardi:2011yz,Gardi:2013ita,Agarwal:2020nyc}, we organise the perturbative expansion in terms of \emph{webs}. These correspond to either diagrams that remain connected upon removing the Wilson lines, or to special linear combinations of diagrams which together give rise to connected colour factors.}. We proceed by reviewing the result and then summarise the key features of the new computation strategy.

\section{The result}
To present the result, we manifest the antisymmetry of ${\cal F}_{1,3}^{(3)}$ in analogy with Eq.~(\ref{Fmassless}), 
\begin{align}
\label{calF_to_F}
\begin{split}
    {\cal F}_{1,3}^{(3)}(r_{ijQ},r_{jkQ},r_{ikQ})=\,\,
    &\\&\hspace{-20mm}
    16\left[F_{1,3}(x,1-\bar{z},1-z)-F_{1,3}\left(x,z,\bar{z}\right)\right],
    \end{split}
\end{align}
where the variables $\{x,z,\bar{z}\}$ are defined by 
\begin{align}
\label{newVar}
\begin{split} r_{ijQ}\equiv\,\,&
\frac{\beta_Q^2\beta_i\cdot\beta_j}{2\beta_i\cdot\beta_Q \beta_j\cdot\beta_Q}
=\frac{x(x+z-\bar{z})}{(1-z)(1-\bar{z})},
   \\
    r_{jkQ}\equiv\,\,&
\frac{\beta_Q^2\beta_j\cdot\beta_k}{2\beta_j\cdot\beta_Q \beta_k\cdot\beta_Q}  =
    \frac{x(x+z-\bar{z})}{z\bar{z}},
    \\
    r_{ikQ}\equiv\,\,&
\frac{\beta_Q^2\beta_i\cdot\beta_k}{2\beta_i\cdot\beta_Q \beta_k\cdot\beta_Q}
    =
    \frac{x(x+z-\bar{z})}{(1-z)(1-\bar{z})z\bar{z}}\,.
\end{split}
\end{align}
 We provide the expression of $F_{1,3}$, written in terms of uniform weight-five GPLs of $\{x,z,\bar{z}\}$,  in~\cite{MathematicaNotebook}. Note that $F_{1,3}$ is invariant under the following Galois symmetries,
\begin{align}
\label{GaloisSymm}
\begin{split}
F_{1,3}(x,z,\bar{z})=\,\,&F_{1,3}(-x,\bar{z},z)=F_{1,3}(-x-z+\bar{z},z,\bar{z}),
    \end{split}
\end{align}
related to flipping the sign of the square roots that appear upon solving~\eqref{newVar} for $\{x,z,\bar{z}\}$. The full one-mass quadrupole in~Eq.~(\ref{cstructure}) is obtained by performing the three permutations on $F_{1,3}$ of Eq.~(\ref{calF_to_F}), as prescribed by~$P$~\footnote{The permutation relations written in terms of $\{x,z,\bar{z}\}$ are given in the supplemental material.}.

In the new variables, the alphabet of ${\cal F}_{1,3}^{(3)}$ is 
\begin{align}
\label{alphabet}
\begin{split}
    &\{\omega_i\}^{\text{phy}}=\,\,\{z,\bar{z},1-z,1-\bar{z}\}
    \cup\{x,\bar{z}-z-x\}\cup\{x+z,
   \\&\,\, \bar{z}-x,
 1-z-x,1-\bar{z}+x,z+x-z\bar{z},\bar{z}-x-z\bar{z}\}\,.
\end{split}
\end{align}
These 12 letters represents physical collinear singularities,  $\beta_u\cdot \beta_v\to 0$. The first group of four letters corresponds
to $r_{uvQ}\to \infty$ and coincides with the full alphabet of~${\cal F}_{0,4}^{(3)}$ in  Eq.~(\ref{Fmassless}). 
The next group of two letters corresponds to $r_{uvQ}\to 0$, while the final group of six letters to $r_{uvQ}\to 1$~\footnote{It is convenient to interpret these limits in the $\beta_Q$ rest frame. There 
$ r_{uv Q}=\frac{1}{2}\left(1-\cos\theta_{uv}\right),
$
where $\theta_{uv}$ is the spatial angle between particles  $u$ and~$v$. The singularities at $r_{uv Q}\to \{0,1\}$ then correspond the collinear ($\theta_{uv}\rightarrow 0$) and the back-to-back ($\theta_{uv}\rightarrow \pi$) configurations, respectively.}. The first symbol entry of ${\cal F}_{1,3}^{(3)}$, associated with its channel discontinuities, is more restricted and only consists of the three ratios $r_{uvQ}$, with branch cuts for $r_{uvQ}>0$.

The collinear limits of ${\cal F}_{1,3}^{(3)}$ are consistent with the  collinear factorization constraints~\cite{Liu:2022elt,Duhr:2025cye}, yielding 
\begin{align}
\label{coll}
     \begin{split}
   {\cal F}_{1,3}^{(3)}(0,r,r)=-{\cal F}_{1,3}^{(3)}(r,0,r)= 2{\cal F}_{0,3}^{(3)},
     \end{split}
     \end{align}
or equivalently, for $F_{1,3}(x,z,\bar{z})$~\footnote{Describing the approach to the collinear limit requires a different parametrization of the kinematics, given in Eq.~(\ref 
{CollinearLImit_parametrization}) in the supplemental material.},
\begin{align}
\label{collF}
\begin{split}
    F_{1,3}(0,1,1)
    =4C\,,
    \quad
     F_{1,3}(0,0,0)
     =0\,.
     \end{split}
 \end{align}
The lightcone  limit $\beta_Q^2\to 0$~is obtained by taking $x\to 0$,  
\begin{equation}
\label{lightlike_limit}
F_{1,3}(0,z,\bar{z})=F_{0,4}(z,\bar{z})\,,
\end{equation}
recovering~the known result~\cite{Almelid:2015jia} of~Eq.~(\ref{Fmassless}). Eq.~(\ref{lightlike_limit}) also represents the triple collinear limit, $\beta_i||\beta_j||\beta_k$~\cite{Duhr:2025cye,Gardi:2025ule}.

\section{Computation strategy}
We start from the correlator~\footnote{We stress that the four velocities are independent: there may be additional non-coloured particles emanating from the hard vertex where the four lines meet.} of four timelike Wilson lines ($\beta_U^2>0$), 
\begin{align}
\label{def_correlator}\left<\Phi_{\beta_I}^{(m)}\Phi_{\beta_J}^{(m)}\Phi_{\beta_K}^{(m)}\Phi_{\beta_Q}^{(m)}\right>\, =\exp\left[\sum_{n}\left(\frac{\alpha_s}{4\pi}\right)^n w^{(n)}\right],
\end{align}
which is multiplicatively renormalizable. Specifically, the renormalization of the multi-Wilson-line vertex involves the very same quadrupole terms appearing in 
${\mathbf \Gamma}$ of Eq.~(\ref{cstructure}), barring the  fact that in the latter three or four velocities are expanded near the lightcone.
As indicated by the superscript~$(m)$, the correlator is regularised in the infrared. While the precise regularization prescription does not affect the ultraviolet pole we aim to extract, we follow previous work~\cite{Gardi:2011yz,Almelid:2015jia}, and introduce 
exponential suppression of the gluon coupling to the Wilson line at large distances, inversely proportional to~$m\sqrt{\beta_U^2}$. Equivalently, in momentum space, the regulator induces a $m\sqrt{\beta_U^2}$ shift in the denominator of the eikonal propagator. The dependence of the regulator on the Wilson-line squared velocity ensures that the correlator remains invariant under the rescaling of each velocity.

Having defined the regularised correlator (\ref{def_correlator}), we follow the strategy described in Ref.~\cite{Gardi:2025ule}, and use the Method of Regions (MoR)~\cite{Beneke:1997zp,Smirnov:2002pj,Pak:2010pt,Jantzen:2011nz,Jantzen:2012mw,Semenova:2018cwy,Ananthanarayan:2018tog,Heinrich:2023til,Borowka:2017idc,Gardi:2022khw,Ma:2023hrt,Chen:2024xwt} to perform an asymptotic lightcone expansion in  
\begin{align}
\label{virtualities_expansion}
    \beta_I^2\sim \lambda_i\,,
    \quad\quad
     \beta_J^2\sim \lambda_j\,,
     \quad\quad
      \beta_K^2\sim \lambda_k\,.
\end{align}
We stress that the naive limit -- dubbed the ``hard region'' -- corresponding to taking the squared velocities in~(\ref{virtualities_expansion}) to zero in the integrand, amounts to losing infrared regularization along the lightlike lines, therefore generating collinear singularities in addition to the ultraviolet ones. This leads to higher-order poles in $\epsilon$, which cannot be linked with the renormalization of the multi-Wilson-line vertex.
However, the full asymptotic expansion obtained by the MoR restores the multiplicatively-renormalizable correlator~\cite{Gardi:2025ule}, replacing the higher-order~$\epsilon$ poles in each web with logarithms of the expansion parameters in (\ref{virtualities_expansion}). These logarithms ultimately cancel in the correlator in the sum of all webs, where one can read off the soft anomalous dimension. This strategy maximises the simplification afforded by the lightlike limit, as each region integral is manifestly free any dependence on the small squared velocities. 

Although the correlator contains non-planar diagrams, there is a Euclidean regime where $\beta_{U}\cdot\beta_{V}<0$ for any pair of velocities $U,V\in\{I,J,K,Q\}$.
Upon performing the asymptotic expansion in this  kinematic regime, we guarantee that all the monomials in the Symanzik graph polynomials in any integral are non-negative, and hence there are no hidden regions~\cite{Gardi:2024axt,Ma:2025emu}. Thus, the complete set of regions can be determined using the facets of the Newton polytope of the graph polynomials~\cite{Pak:2010pt,Jantzen:2012mw,Semenova:2018cwy,Gardi:2022khw,Ananthanarayan:2018tog,Heinrich:2023til,Borowka:2017idc,Chen:2024xwt}. In this work we used the packages {\tt pySecDec}~\cite{Heinrich:2021dbf} and {\tt AmpRed}~\cite{Chen:2024xwt} to determine the region vectors.

Performing the perturbative expansion of the correlator~(\ref{def_correlator}), one finds that there are four types of four-line webs (see Fig.~\ref{fig:threeloop4linewebs})~\cite{Frenkel:1984pz,Gatheral:1983cz,Sterman:1981jc,Mitov:2010rp,Gardi:2010rn,Gardi:2011wa,Gardi:2011yz,Gardi:2013ita,Almelid:2015jia,Almelid:2016lrq} that contribute to the colour quadrupole structure $\mathbf{T}_{uv;w Q}$ at three loops~\cite{Almelid:2015jia},
\begin{align}
\label{FourLineWebs}
\left(\frac{\alpha_s}{4\pi}\right)^3w^{(3)}= W_{1111}+W_{1112}+W_{1122}+W_{1113}+\cdots\,,
\end{align}
where the subscripts on $W$ count the number of attachments to each of the four Wilson lines. Each $W$ term  represents the sum of all webs of a given type. The ellipsis represent webs that span fewer lines.

\section{Connected webs}
$W_{1111}$ represents the sum of all connected diagrams spanning four lines. These feature either a single four gluon vertex or two three-gluon vertices (see Fig.~\ref{fig:threeloop4linewebs}).
These diagrams are special: they are the only ones that give rise to integrals depending on all three rescaling-invariant ratios of Eq.~(\ref{newVar}). These are thus by far the most complicated integrals contributing to ${\cal F}_{1,3}^{(3)}$. We therefore focus our description of the calculation mainly on these webs~\footnote{A complete description of the computation will be reported elsewhere~\cite{GZ-TBP}.}. 

The structure of $W_{1111}$ is~\footnote{The subscript of the kinematic function ${\cal Y}$ in (\ref{W1111calY}) indicates that the two velocities on either side of the semicolon are directly connected by a three-~or four-~gluon vertex. By Bose symmetry these functions are antisymmetric, e.g. ${\cal Y}_{(IJ;KQ)}=
-{\cal Y}_{(JI;KQ)}=-{\cal Y}_{(IJ;QK)}$.}
\begin{eqnarray}
\label{W1111calY}
W_{1111}\,&=&\,\textbf{T}_I^a\textbf{T}_J^b\textbf{T}_K^c\textbf{T}_Q^d\big[f^{abe}f^{cde}{\cal Y}_{(IJ;KQ)}(\{\alpha_{VU}\})
\nonumber \\&&\hspace*{10pt}
+f^{ace}f^{bde}{\cal Y}_{(IK;JQ)} (\{\alpha_{VU}\})
\\&& \hspace*{10pt}
+f^{ade}f^{bce}{\cal Y}_{(JK;IQ)}(\{\alpha_{VU}\})\big]\,.\nonumber
\end{eqnarray}
The three channels are related by permutations, so let us now consider ${\cal Y}_{(IJ;KQ)}$.  Relating the correlator~(\ref{def_correlator}) to the anomalous dimension~(\ref{cstructure}) requires to compute ${\cal Y}$ with four timelike Wilson lines, so prior to the expansion, the kinematic dependence is via the set $\{\alpha_{VU}\}$ of \emph{six} rescaling-invariant variables,
\begin{align}
\label{alphaDef}
    \alpha_{VU}+\frac{1}{\alpha_{VU}}\equiv-\frac{2\beta_V\cdot\beta_U}{\sqrt{\beta_V^2\beta_U^2}}\,.
\end{align}
Being connected, these diagrams features a single ultraviolet pole in $\epsilon$,
\begin{align}
\label{Laurent}
    {\cal Y}_{(IJ;KQ)}(\{\alpha_{VU}\})=\sum_{l=-1}^{\infty}{\cal Y}_{(IJ;KQ)}^{(l)}(\{\alpha_{VU}\})\,\epsilon^l\,.
\end{align}
The connected nature of the diagrams also guarantees that it suffices to regularize the infrared on a single Wilson line~\cite{Henn:2023pqn}. We therefore drop the regulator $m$ on the lines $I$, $J$ and~$K$,  while keeping it on line $Q$, without affecting the ultraviolet $1/\epsilon$ pole we are interested in. In this setup the expansions in the three parameters of Eq.~(\ref{virtualities_expansion}) commute. 

\begin{widetext}
The asymptotic expansion of the Laurent coefficients in Eq.~(\ref{Laurent}) takes the form a log-power expansion. Specifically, at the leading order in~$\epsilon$, ${\cal O}(\epsilon^{-1})$, we have
\begin{align}
\label{CalYExpected}
\begin{split}
   &{\cal T}_{\lambda_i,\lambda_j,\lambda_k}\left[ {\cal Y}_{(IJ;KQ)}^{(-1)}(\{\alpha_{VU}\})\right]= 
{\cal Y}_{(ij;kQ)}^{(-1)}(r_{ikQ},r_{jkQ},r_{ijQ}; \lambda_i, \lambda_j, \lambda_k) 
   \,+\, {\cal O}(\lambda)\,,
   \end{split}
\end{align}
where ${\cal Y}_{(ij;kQ)}^{(-1)}$ on the right-hand side 
depends on the kinematics via the ratios 
$r_{uvQ}$ as well as the three logarithms of the expansion parameters $\lambda_u$ of Eq.~(\ref{virtualities_expansion}).

We compute the asymptotic lightcone expansion using the MoR, which 
yields a sum over region integrals:
\begin{eqnarray}
\label{CalYExp}
 & {\cal T}_{\lambda_i,\lambda_j,\lambda_k}\left[ {\cal Y}_{(IJ;KQ)}(\{\alpha_{VU}\})\right]= \displaystyle{\sum_{[n_i,n_j,n_k]}}
\lambda_i^{n_i\epsilon}
\lambda_j^{n_j\epsilon}
\lambda_k^{n_k\epsilon} 
\,\, {\cal Y}_{(ij;kQ)}^{[n_i,n_j,n_k]}(r_{ikQ},r_{jkQ},r_{ijQ})\,,
\end{eqnarray}
where
we neglected all subleading powers in the expansion parameters~$\lambda_i$, $\lambda_j$ and $\lambda_k$. Here the regions are designated by $[n_i,n_j,n_k]$, identifying their scaling law~$\lambda_u^{n_u\epsilon}$ for $u\in \{i,j,k\}$, where $n_u$ are integers. 
In each region integral we have scaled out the non-analytic dependence on the expansion parameters. Crucially, any given region integral ${\cal Y}_{(ij;kQ)}^{[n_i,n_j,n_k]}$~depends \emph{only} on the three $r_{uvQ}$ ratios, and not on the expansion parameters. Individual region integrals feature higher-order poles in $\epsilon$. These cancel in the sum of regions, leaving a trace in ${\cal Y}^{(-1)}_{(ij;kQ)}$ as  logarithms of the three expansion parameters, in line with Eq.~(\ref{CalYExpected}).  

Applying the MoR to $W_{1111}$, we get $56$ regions for each of the three channels. Most of the region integrals are simple to evaluate; the most complicated $8$ region integrals depend on all three $r_{uvQ}$ ratios via polylogarithms. 
These are the hard region, where the three Wilson lines are taken to be strictly lightlike, and 7  other regions which feature collinear modes:
\begin{align}
\label{MoRWeb111}
\begin{split}
   {\cal T}_{\lambda_i,\lambda_j,\lambda_k}\left[W_{1111}\right]=&\vcenter{\hbox{\includegraphics[width=2cm]{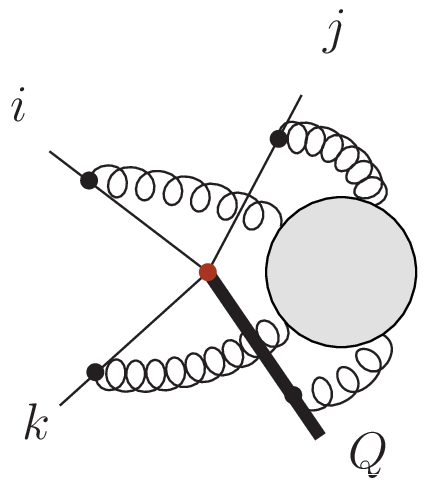}}}=\vcenter{\hbox{\includegraphics[width=2cm]{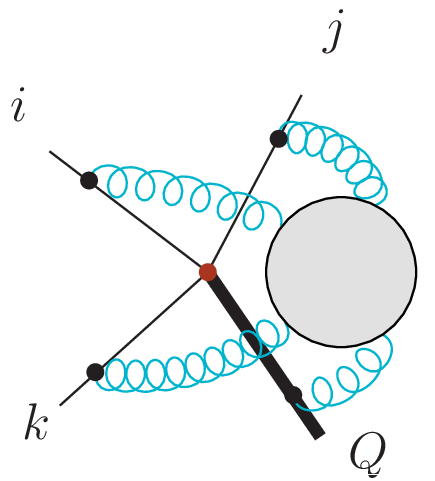}}}+\lambda_i^{-\epsilon}\vcenter{\hbox{\includegraphics[width=2cm]{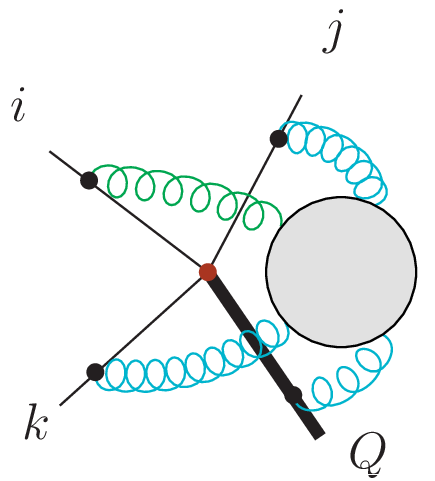}}}+\lambda_j^{-\epsilon}\vcenter{\hbox{\includegraphics[width=2cm]{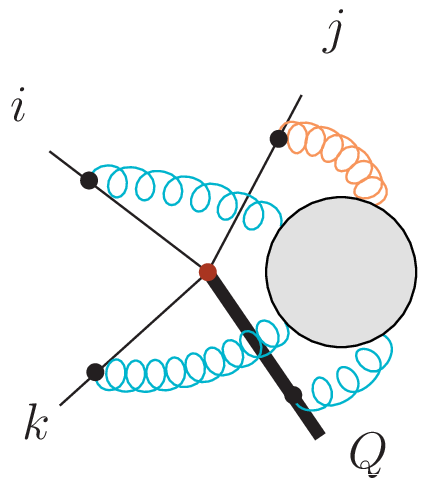}}}+\lambda_k^{-\epsilon}\vcenter{\hbox{\includegraphics[width=2cm]{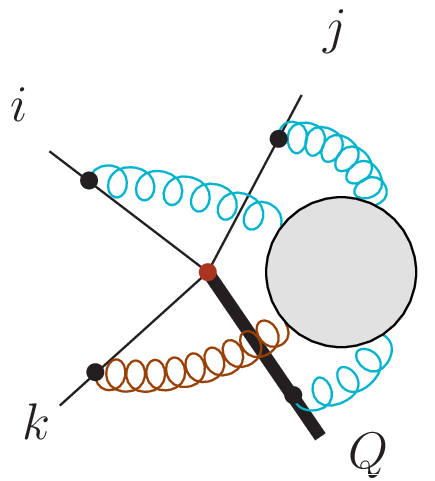}}}
\\&\hspace{-20mm}+\lambda_i^{-\epsilon}\lambda_j^{-\epsilon}\vcenter{\hbox{\includegraphics[width=2cm]{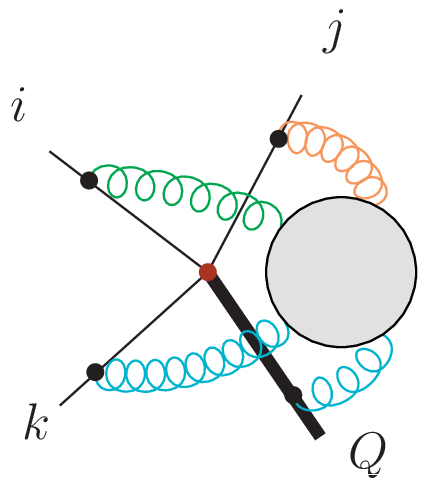}}}+\lambda_i^{-\epsilon}\lambda_k^{-\epsilon}\vcenter{\hbox{\includegraphics[width=2cm]{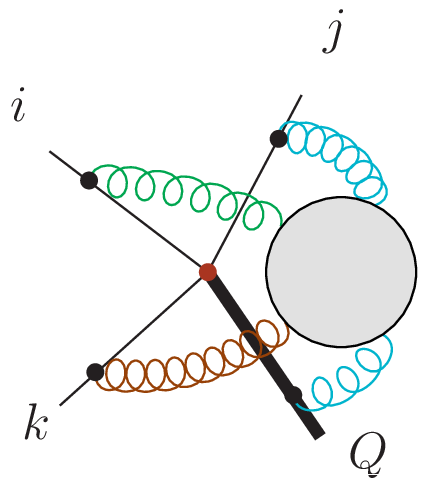}}}+\lambda_j^{-\epsilon}\lambda_k^{-\epsilon}\vcenter{\hbox{\includegraphics[width=2cm]{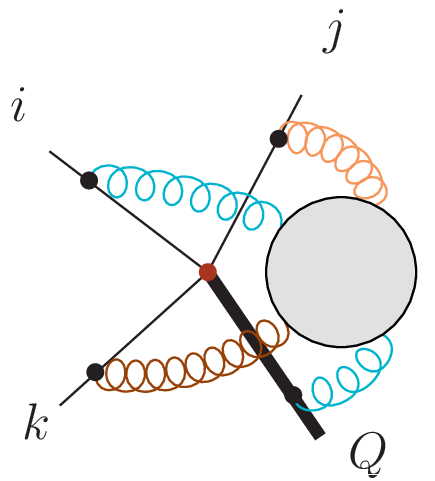}}}+\lambda_i^{-\epsilon}\lambda_j^{-\epsilon}\lambda_k^{-\epsilon}\vcenter{\hbox{\includegraphics[width=2cm]{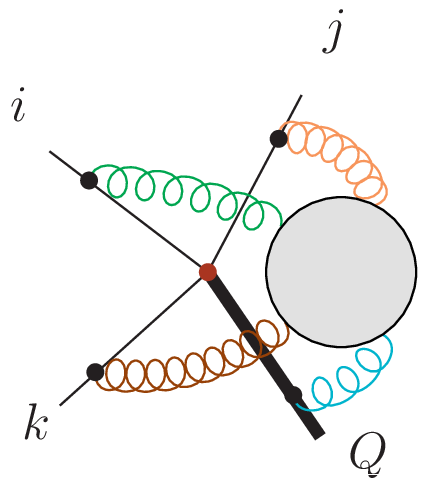}}}+\ldots\,.
 \end{split}
 \end{align}
  \end{widetext}
Here the gluons are colour-coded according to the loop-momentum modes~\cite{Gardi:2025ule}: the blue gluons are hard, while the green, orange and brown gluons are collinear to the $i$, $j$ and $k$ directions, respectively.  The blob represents all the possible ways in which the four gluons connect through a single four-gluon vertex or two three-gluon vertices.

We use the method of differential equation~\cite{Henn:2013pwa} to compute the region integrals.  Using the packages {\tt AmpRed}~\cite{Chen:2024xwt}, {\tt Kira}~\cite{Maierhofer:2017gsa,Klappert:2020nbg,Lange:2025fba} and {\tt LiteRed}~\cite{Lee:2012cn,Lee:2013mka}, we derive the differential equation for a subset of the regions of one channel, obtaining the rest by permutations. In addition to the singularities at $r_{uvQ}\rightarrow \{0,1,\infty\}$, represented by the alphabet (\ref{alphabet}), the differential equations display singularities also at~${\textcolor{blue}{\Delta_1}}\rightarrow 0$ and ${\textcolor{blue}{\Delta_2}}\rightarrow 0$, where
 \begin{align}
 \label{NewSing}
 \begin{split}
     \textcolor{blue}{\Delta_1}=\,\,\Delta(r_{ijQ},r_{ikQ},r_{jkQ}),
\quad    \textcolor{blue}{\Delta_2}=\,\,\textcolor{blue}{\Delta_1}+4r_{ijQ}r_{jkQ}r_{ikQ}\,, 
\end{split}
 \end{align}
where $\Delta$ is the K\"{a}ll\'{e}n function. $\textcolor{blue}{\Delta_1}$ and $\textcolor{blue}{\Delta_2}$ are quadratic and cubic polynomias in the $r_{ijQ}$ variables, respectively. 
In the parametrization (\ref{newVar}) this corresponds to the following extended alphabet
\begin{align}
\label{extended_alphabet}
\begin{split}
    \{\omega_i\}=
    \{\omega_i\}^{\text{phy}}\cup
\{\textcolor{blue}{z-\bar{z}}, \textcolor{blue}{2x+z-\bar{z}}\}\,,
\end{split}
\end{align}
where the last two letters originate exclusively in $\textcolor{blue}{\Delta_1}$ and~$\textcolor{blue}{\Delta_2}$~\footnote
{Rationalising the extended alphabet was the original incentive to introduce the parametrization in Eq.~\eqref{newVar}. In obtaining this parametrization we used the package {\tt RationalizeRoots}~\cite{Besier:2019kco}.}.
These additional letters only appear in $W_{1111}$, not in any other contribution to the correlator.   

In order to solve the differential equation, we bring it to an $\epsilon$-factorized form~\cite{Henn:2013pwa}. To this end we first identify one uniformly transcendental integral. 
Based on the experience from 
calculating the fully  lightlike limit~\cite{Almelid:2015jia,Almelid:2016lrq}, one expects that this would be a property of the specific integral entering the anomalous dimension, i.e. ${\cal T}_{\lambda_i,\lambda_j,\lambda_k}\left[{\cal Y}_{(IJ;KQ)}\right]$.
This is indeed so. Moreover,  each region (characterized by distinct scaling 
$\lambda_{i}^{n_i\epsilon}\lambda_{j}^{n_j\epsilon}
\lambda_{j}^{n_j\epsilon}$) 
is individually uniformly transcendental. Based on this, using the package {\tt Initial}~\cite{Dlapa:2020cwj}, we obtained a canonical form of the differential equations of each region. The general solution is then obtained, order-by-order in $\epsilon$ in terms of generalized polylogarithms using {\tt PolyLogTools} \cite{Duhr:2019tlz}. 

Important simplifications occur in~$W_{1111}$ upon summing all region integrals. As discussed in the  supplemental material, the result is compatible with a  range of physical constraints, including Bose-symmetry, the first-entry condition, a single ultraviolet $1/\epsilon$ pole, the known $\beta_Q^2\to 0$ limit, obtained in the computation of the  connected webs in Ref.~\cite{Almelid:2015jia,Almelid:2016lrq}, and the known collinear limit for the soft anomalous dimension,  determined by the strict collinear factorization requirement~\cite{Liu:2022elt,Duhr:2025cye}.   These constraints, put together, fully fix the boundary data, yielding a unique result for $W_{1111}$. We stress that in contrast to individual region integrals, this result is free of higher-order poles in~$\epsilon$, and the leading order ${\cal Y}^{(-1)}_{(ij;kQ)}$ is free of singularities at $\textcolor{blue}{\Delta_1}$ and~$\textcolor{blue}{\Delta_2}$. It is a uniform weight-five function with GPLs of the physical alphabet~(\ref{alphabet}) and logarithms of the expansion parameters.

\section{Other webs}
The remaining webs~\footnote{Note that non-connected webs have subdivergences. In the process of renormalizing the multi-Wilson-line vertex, these are removed by commutators of lower-loop webs. The three-loop soft anomalous dimension may be written in terms of $w^{(n)}$ as in Eq.~(2.15c) in Ref.~\cite{Gardi:2011yz}.} in Eq.~\eqref{FourLineWebs}, do not involve functions of all three kinematic variables and are therefore simpler.
The most complicated of these is $W_{1112}$ (see Fig.~\ref{fig:threeloop4linewebs}), which has one three-gluon vertex. The calculation strategy is similar to that of the connected webs, so here we describe it briefly, referring to the key differences. 
In contrast to $W_{1111}$, the $W_{1112}$ integrals feature double poles in $\epsilon$, making them sensitive to the regularization. It is therefore convenient to identify the three expansion parameters $\lambda_i=\lambda_j=\lambda_k\equiv\lambda$ before expanding. In the calculation of this web, we use the package {\tt CANONICA}~\cite{Meyer:2017joq} to obtain the canonical form of the differential equation of each region. In these webs, the two singularities $\textcolor{blue}{\Delta_i}$ of Eq.~\eqref{NewSing} do not appear in the differential equations, but instead we find new singularities, $1+r_{uvQ}$ and $r_{uvQ}+r_{uwQ}$. Eventually, these do not contribute to the single pole in $\epsilon$. 

The remaining two webs in Eq.~\eqref{FourLineWebs}, $W_{1122}$ and~$W_{1113}$ (see Fig.~\ref{fig:threeloop4linewebs}), are multi-gluon-exchange webs; these have been computed with all four velocities timelike in terms of GPLs~\cite{Gardi:2013saa,Falcioni:2014pka}. It is therefore straightforward to expand these results according to Eq.~(\ref{virtualities_expansion}).

In addition to the four-line webs in Eq.~\eqref{FourLineWebs}, three-~and two-line webs also contribute to $F_{1,3}$ upon applying colour conservation~\cite{Almelid:2015jia,Almelid:2016lrq}~\footnote{See Eqs.~(10) and (11) in~\cite{Almelid:2015jia}.}. These contributions can be either obtained from known results~\cite{Almelid:2016lrq}, or deduced using Eq.~\eqref{coll}. In the sum of the four-line webs and those with fewer lines, we find that the $\lambda$ dependence entirely cancels out, and $F_{1,3}$ only depends on the variables $\{x,z,\bar{z}\}$, as expected. 

\section{Conclusions}
In this paper we computed the three-loop soft anomalous dimension governing the singularities of QCD amplitudes with one heavy particle and any number of massless ones, summarised in Eq.~(\ref{cstructure}). All ingredients except the most complicated one -- the quadrupole function ${\cal F}_{1,3}^{(3)}(r_{ijQ},r_{jkQ},r_{ikQ})$ correlating the heavy particle with three massless ones -- have been determined before, so powerful consistency checks are available. 
In particular, the lightlike limit of ${\cal F}_{1,3}^{(3)}$ reproduces the well-known massless quadrupole function ${\cal F}_{0,4}^{(3)}(\rho_{ijkq},\rho_{kjiq})$~\cite{Almelid:2015jia}, as stated in Eq.~\eqref{lightlike_limit}. The latter also describes the three-particle collinear limit~\cite{Duhr:2025cye}. Furthermore, the two-particle collinear limits of ${\cal F}_{1,3}^{(3)}$ are independent of any kinematic variables, and are given by Eq.~\eqref{coll}, in line with the constraints deduced from strict collinear factorization~\cite{Liu:2022elt,Duhr:2025cye}.

Beyond said kinematic limits, our result for ${\cal F}_{1,3}^{(3)}$ admits a highly-constrained analytic structure, in line with expectations: the symbol alphabet~(\ref{alphabet}) directly relates to physical collinear singularities corresponding to $r_{uvQ}\to \{0,1,\infty\}$. The only discontinuity accessible from the first sheet is the branch cut for $r_{uvQ}>0$, a manifestation of the first-entry condition. The result respects the two Galois symmetries in Eq.~\eqref{GaloisSymm}. Finally, the three channels related by permuting lightlike velocities are all real in the Euclidean regime.

To obtain the new result for ${\cal F}_{1,3}^{(3)}$ we used a novel strategy, which retains the advantages of  multiplicatively renormalizing a manifestly process-independent, gauge-invariant and rescaling-invariant correlator of timelike semi-infinite Wilson lines, while taking full advantage of the simplification afforded by the lightlike limit through the use of the Method of Regions~\cite{Gardi:2025ule}. The most striking aspect of this
simplification is the fact that each of the region integrals ${\cal Y}_{(ij;kQ)}^{[n_i,n_j,n_k]}$ we computed depends on just three rescaling-invariant ratios, the variables that remain finite in the limit, rather than the full set of six independent variables on which the correlator depends. This strategy opens the way to compute the soft anomalous dimension at higher loop orders and for amplitudes involving more heavy particles at three loops.

\section*{Acknowledgments}

We are grateful to Wen Chen, Claude Duhr, Stephen Jones, Gregory Korchemsky, Yao~Ma, Andrew McLeod and Kai~Yan for helpful discussions. 
ZZ is supported by the China Scholarship Council PhD programme. 
EG is supported by the STFC Consolidated Grant \emph{Particle Physics at the Higgs Centre}. 
For the purpose of open access, the authors have applied a Creative Commons Attribution (CC BY) licence to any Author Accepted Manuscript version arising from this submission.

\appendix

\bibliography{biblio}
\bibliographystyle{apsrev4-2}


\onecolumngrid
\newpage
\appendix

\section*{Supplemental material}
\subsection{One-mass soft anomalous dimension to three loops}
\label{sec:ThreeLoopResults}

\subsubsection{Anomalous dimensions contributing to the dipole component}

Eq.~\eqref{cstructure} summarises the one-mass soft anomalous dimension for any amplitude involving one heavy particle $Q$ and any number of massless ones.
Here we collect the known results to three loops, for each of the components entering Eq.~\eqref{cstructure}. Firstly, we define the perturbative expansion of $\gamma_{\text{cusp}}$, $\Omega_{Q}$ and $\gamma_{i}$,
\begin{align}
    \begin{split}
\gamma_{\text{cusp}}=\sum_{n=1}^{\infty}\gamma_{\text{cusp}}^{(n)}\left(\frac{\alpha_s}{4\pi}\right)^n\,,
    \quad
    \Omega_{Q}=\sum_{n=1}^{\infty}\Omega_{Q}^{(n)}\left(\frac{\alpha_s}{4\pi}\right)^n\,,
    \quad
    \gamma_{i}=\sum_{n=1}^{\infty}\gamma_{i}^{(n)}\left(\frac{\alpha_s}{4\pi}\right)^n\,.
    \end{split}
\end{align}
The coefficients $\gamma_{\text{cusp}}^{(n)}$ to three loops~\cite{Moch:2004pa,Korchemsky:1987wg} are given by 
\begin{subequations}
\begin{align}
   \begin{split}
       \gamma_{\text{cusp}}^{(1)}=\,\,&4\,,
   \end{split} 
   \\
   \begin{split}
       \gamma_{\text{cusp}}^{(2)}=\,\,&\left(\frac{268}{9}-\frac{4}{3}\pi^2\right)C_A-\frac{80}{9}n_fT_F\,,
   \end{split}
   \\
   \begin{split}
\gamma_{\text{cusp}}^{(3)}=\,\,&\left(\frac{490}{3}-\frac{536}{27}\pi^2+\frac{88}{3}\zeta(3)+\frac{44}{45}\pi^4\right)C_A^2
    +\left(-\frac{1672}{27}+\frac{160}{27}\pi^2-\frac{224}{3}\zeta(3)\right)C_An_fT_F
\\&+\left(-\frac{220}{3}+64\zeta(3)\right)C_Fn_fT_F-\frac{64}{27}\left(n_fT_F\right)^2\,.
\end{split}
\end{align}
\end{subequations}
Then, $\Omega_{Q}^{(n)}$ to three loops~\cite{Becher:2009kw,Bruser:2019yjk} are
\begin{subequations}
\begin{align}
   \begin{split}
       \Omega_{Q}^{(1)}=\,\,&-2C_Q\,,
   \end{split} 
   \\
   \begin{split}
       \Omega_{Q}^{(2)}=\,\,&C_Q\left[\left(-\frac{98}{9}+\frac{2}{3}\pi^2-4\zeta(3)\right)C_A+\frac{40}{9}n_fT_F\right]\,,
   \end{split}
   \\
   \begin{split}
 \Omega_{Q}^{(3)}=\,\,&C_Q\bigg[\bigg(-\frac{343}{9}+\frac{304}{27}\pi^2-\frac{740}{9}\zeta(3)-\frac{22}{45}\pi^4
 -\frac{4}{3}\pi^2\zeta(3)+36\zeta(5)\bigg)C_A^2
    \\&+\left(\frac{356}{27}-\frac{80}{27}\pi^2+\frac{496}{9}\zeta(3)\right)C_An_fT_F+\left(\frac{110}{3}-32\zeta(3)\right)C_Fn_fT_F+\frac{32}{27}\left(n_fT_F\right)^2\bigg]\,.
\end{split}
\end{align}
\end{subequations}
Finally, $\gamma_{i}^{(n)}$ to three loops~\cite{Becher:2006mr,Baikov:2009bg} are
\begin{subequations}
\begin{align}
   \begin{split}
       \gamma_{i}^{(1)}=\,\,&-3C_i\,,
   \end{split} 
   \\
   \begin{split}
\gamma_{i}^{(2)}=\,\,&C_i\bigg[\left(-\frac{961}{54}-\frac{11}{6}\pi^2+26\zeta(3)\right)C_A+\left(-\frac{3}{2}+2\pi^2-24\zeta(3)\right)C_F+\left(\frac{130}{27}+\frac{2}{3}\pi^2\right)n_fT_F\bigg]\,,
   \end{split}
   \\
   \begin{split}
 \gamma_{i}^{(3)}=\,\,&C_i\bigg[\bigg(-\frac{139345}{2916}-\frac{7163}{486}\pi^2+\frac{3526}{9}\zeta(3)-\frac{83}{90}\pi^4-\frac{44}{9}\pi^2\zeta(3)-136\zeta(5)\bigg)C_A^2
    \\&+\bigg(-\frac{151}{4}+\frac{205}{9}\pi^2-\frac{844}{3}\zeta(3)
    +\frac{247}{135}\pi^4-\frac{8}{3}\pi^2\zeta(3)-120\zeta(5)\bigg)C_AC_F
\\&+\bigg(-\frac{29}{2}-3\pi^2-68\zeta(3)
-\frac{8}{5}\pi^4+\frac{16}{3}\pi^2\zeta(3)+240\zeta(5)\bigg)C_F^2
\\&+\bigg(-\frac{17318}{729}+\frac{2594}{243}\pi^2
-\frac{1928}{27}\zeta(3)+\frac{22}{45}\pi^4\bigg)C_An_fT_F
\\&+\bigg(\frac{2953}{27}-\frac{26}{9}\pi^2+\frac{512}{9}\zeta(3)-\frac{28}{27}\pi^4\bigg)C_Fn_fT_F
\\&+\bigg(\frac{9668}{729}-\frac{40}{27}\pi^2-\frac{32}{27}\zeta(3)\bigg)\left(n_fT_F\right)^2\bigg]\,.
\end{split}
\end{align}
\end{subequations}
The quadratic Casimir is defined by $\mathbf{T}_i^a\mathbf{T}_i^a\equiv C_i\mathbf{1}$. The subscripts $``A"$ and $``F"$ represent adjoint and fundamental colour representations, respectively. $T_F$ is the normalization of the generators  in fundamental representation, i.e.,  $\text{Tr}\left[\mathbf{T}_F^a\mathbf{T}_F^b\right]\equiv T_F\delta_{ab}$. Finally, $n_f$ counts the number of massless quarks with different flavours.

\subsubsection{The three channels of the quadrupole component}

\begin{figure}[h]
\begin{subfigure}{0.18\linewidth}
    \centering \includegraphics[width=1\linewidth]{figures/3g3g1Tchik.eps}
    \end{subfigure}
    \hfill
    \begin{subfigure}{0.18\linewidth}
    \centering \includegraphics[width=1\linewidth]{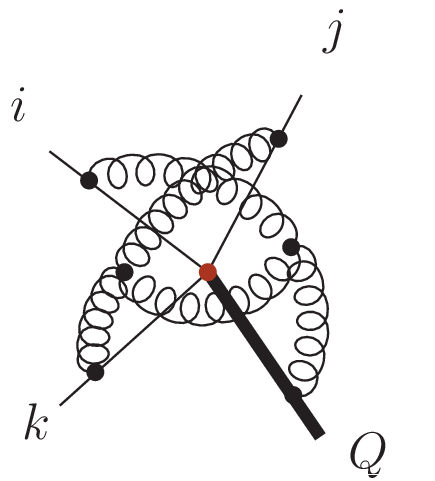}
    \end{subfigure}
    \hfill
     \begin{subfigure}{0.18\linewidth}
    \centering \includegraphics[width=1\linewidth]{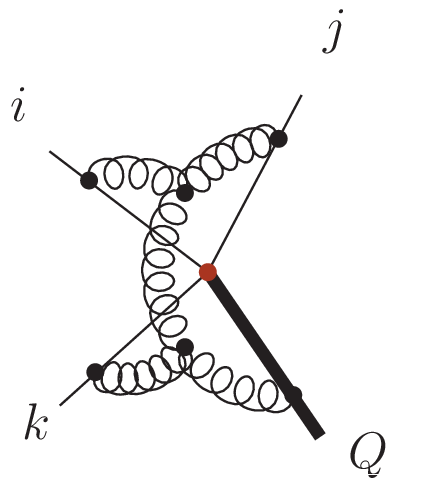}
    \end{subfigure}
     \caption{The diagrams with two three-gluon vertices contributing to the three channels $P$. From left to right, the diagrams are associated with the colour factor $\mathbf{T}_{uv;w Q}$ where the triplet of indices $(u,v;w)$
    is in $P$.}
\label{fig:ThreeChannelsTwo3g}
\end{figure}

The quadrupole terms in Eq.~(\ref{cstructure}) are written as a sum over permutations of the three channels. 
In Fig.~\ref{fig:ThreeChannelsTwo3g} we show the diagrams with two three-gluon vertices defining the three channels.  These three carry the corresponding colour structures 
$\mathbf{T}_{uv;w Q}$ with $(u,v;w)\in P\equiv \{(i,k;j),(j,k;i),(i,j;k)\}$.

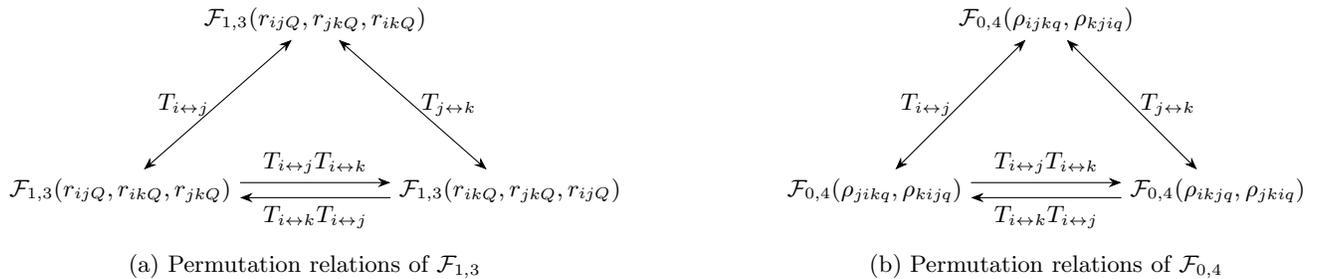
\begin{figure}[bt]
\begin{subfigure}{0.45\linewidth}
\centering
\begin{tikzpicture}[>=Stealth, node distance=2cm, auto]

  \node (A) {${\cal F}_{1,3}(r_{ijQ},r_{ikQ},r_{jkQ})$};
  \node (B) [right=of A] {${\cal F}_{1,3}(r_{ikQ},r_{jkQ},r_{ijQ})$};
  \node (C) [above=of $(A)!0.5!(B)$] {${\cal F}_{1,3}(r_{ijQ},r_{jkQ},r_{ikQ})$};

  \coordinate (Aup) at ($(A.east)+(0,0.1)$);
  \coordinate (Adown) at ($(A.east)+(0,-0.1)$);
  \coordinate (Bup) at ($(B.west)+(0,0.1)$);
  \coordinate (Bdown) at ($(B.west)+(0,-0.1)$);

  \draw[->] (Aup) -- node[above] {$T_{i\leftrightarrow j}T_{i\leftrightarrow k}$} (Bup);
  \draw[->] (Bdown) -- node[below] {$T_{i\leftrightarrow k}T_{i\leftrightarrow j}$} (Adown);

  \draw[<->] (B) -- node[right] {$T_{j\leftrightarrow k}$} (C);
  \draw[<->] (C) -- node[left] {$T_{i\leftrightarrow j}$} (A);

\end{tikzpicture}
\caption{Permutation relations of ${\cal F}_{1,3}$}
\label{Fig:Permu13}
\end{subfigure}
 \hfill
\begin{subfigure}{0.45\linewidth}
\centering
\begin{tikzpicture}[>=Stealth, node distance=2cm, auto]

  \node (A) {${\cal F}_{0,4}(\rho_{jikq},\rho_{kijq})$};
  \node (B) [right=of A] {${\cal F}_{0,4}(\rho_{ikjq},\rho_{jkiq})$};
  \node (C) [above=of $(A)!0.5!(B)$] {${\cal F}_{0,4}(\rho_{ijkq},\rho_{kjiq})$};

  \coordinate (Aup) at ($(A.east)+(0,0.1)$);
  \coordinate (Adown) at ($(A.east)+(0,-0.1)$);
  \coordinate (Bup) at ($(B.west)+(0,0.1)$);
  \coordinate (Bdown) at ($(B.west)+(0,-0.1)$);

  \draw[->] (Aup) -- node[above] {$T_{i\leftrightarrow j}T_{i\leftrightarrow k}$} (Bup);
  \draw[->] (Bdown) -- node[below] {$T_{i\leftrightarrow k}T_{i\leftrightarrow j}$} (Adown);
  
  \draw[<->] (B) -- node[right] {$T_{j\leftrightarrow k}$} (C);
  \draw[<->] (C) -- node[left] {$T_{i\leftrightarrow j}$} (A);

\end{tikzpicture}
\caption{Permutation relations of ${\cal F}_{0,4}$}
\label{Fig:Permu04}
\end{subfigure}
\caption{Permutation relations between the three functions in the sum over $P=\{(i,k;j),(j,k;i),(i,j;k)\}$ in Eq.~\eqref{cstructure}. The two functions at the bottom of the triangle are related by a sequence of two permutations, where the role $T_{i\leftrightarrow j}$ is to generate an extra minus sign.}
\label{Fig:Permu}
\end{figure}

In Eq.~(\ref{cstructure}), for any given subset of massless particles $i$, $j$ and $k$, the 
triplet of indices $(u,v;w)$ is identified with each of the following three permutations $\{(i,k;j),(j,k;i),(i,j;k)\}$.  These permutations apply simultaneously to the colour factor in Eq.~(\ref{Tuvwq}) and to the kinematically-dependent function 
${\cal F}_{1,3}^{(3)}(r_{ijQ},r_{jkQ},r_{ikQ})$
in Eq.~(\ref{calF_to_F}) for the one-mass case, 
\begin{align}
\label{F133appendix}
\begin{split}
    {\cal F}_{1,3}^{(3)}(r_{ijQ},r_{jkQ},r_{ikQ})=\,\,
    16\left[F_{1,3}(x,1-\bar{z},1-z)-F_{1,3}\left(x,z,\bar{z}\right)\right]\,.
    \end{split}
\end{align}
and similarly the function ${\cal F}_{0,4}^{(3)}(\rho_{ijkQ},\rho_{jkiQ})$ in Eq.~(\ref{Fmassless}) for the massless case.
Figure~\ref{Fig:Permu} depicts these   permutations graphically.  Using Eq.~(\ref{newVar}) we can readily translate the permutation into an operation on the functional dependence of the function $F_{1,3}\left(x,z,\bar{z}\right)$ on the variables $\{x, z, \bar{z}\}$ as follows,
    \begin{align}
    \label{PermutationRelations}
\begin{split}
\begin{array}{llll}
 \displaystyle T_{i\leftrightarrow j}:\qquad 
& \displaystyle x\rightarrow \frac{x}{(1-z)(1-\bar{z})},
& \quad  \displaystyle z\rightarrow \frac{\bar{z}}{\bar{z}-1}, 
& \quad  \displaystyle \bar{z}\rightarrow \frac{z}{z-1},
\\ 
\\
\displaystyle T_{j\leftrightarrow k}:\qquad & \displaystyle x\rightarrow \frac{x}{z\bar{z}},
& \quad  \displaystyle z\rightarrow \frac{1}{\bar{z}},
&\quad  \displaystyle \bar{z}\rightarrow \frac{1}{z},
\\
\\
 \displaystyle T_{i\leftrightarrow k}:\qquad  & \displaystyle x\rightarrow x, 
& \quad \displaystyle z\rightarrow 1-\bar{z},
& \quad \displaystyle \bar{z}\rightarrow 1-z.
\end{array}
\end{split}
\end{align}
These operations also apply to $F_{0,4}(z,\bar{z})$ upon ignoring the transformation of $x$. By Bose symmetry, the antisymmetry of the right-hand side of Eq.~(\ref{F133appendix}) upon applying $T_{i\leftrightarrow k}$ is a reflection of the antisymmetry of $\mathbf{T}_{ik;j Q}$, a property of the three-gluon vertex in the leftmost diagram of~Fig.~\ref{fig:ThreeChannelsTwo3g}.

The complete quadrupole contribution in Eq.~(\ref{cstructure}) can be obtained using Eqs.~\eqref{Fmassless},~\eqref{calF_to_F},~\eqref{PermutationRelations} and Fig.~\ref{Fig:Permu}, with $F_{0,4}(z,\bar{z})$ and $F_{1,3}(x, z,\bar{z})$ as input. Explicit expressions for the latter functions, as well as for ${\cal F}_{1,3}^{(3)}$ in the three channels, in terms of GPLs of $\{x,z,\bar{z}\}$, are provided in our \texttt{GitHub} repository~\cite{MathematicaNotebook} in the form of a Mathematica notebook.

Finally, comparing our notation in eq.~\eqref{cstructure} with that in  Ref.~\cite{Liu:2022elt} we find:
\begin{align}
{\cal F}_{0,4}=8F_4, 
\quad
 {\cal F}_{0,3}=2f, 
 \quad
 {\cal F}_{1,3}=2F_{h3}, 
 \quad
 {\cal F}_{1,2}=2F_{h2}.
 \end{align}

\subsection{Boundary conditions of the connected webs}

The connected web $W_{1111}$ of Eq.~(\ref{W1111calY}) consists of three channels, related by permutations. 
We choose to consider $ {\cal Y}_{(ij;kQ)}$ (the other two channels in $W_{1111}$ are obtained by permutations given in Eq.~\eqref{PermutationRelations}). We compute this function as Laurent series in $\epsilon$ by solving differential equations order by order. While the differential equations are set up separately for each region integral, we sum up the general solutions before imposing the boundary conditions. Once the boundary conditions are imposed, this sum represents the true asymptotic expansion of Eq.~(\ref{W1111calY}), defined with four timelike Wilson lines.

The general solution contains a large number of boundary-value parameters to be fixed. Here we explain how we determine these for $ {\cal Y}_{(ij;kQ)}$. The constraints we impose on $ {\cal Y}_{(ij;kQ)}$ include Bose symmetry, single $1/\epsilon$ pole, the lightcone expansion ($\beta_Q^2\to 0$), the collinear expansion ($\beta_j||\beta_k$), and the first-entry condition. The fact that these physical conditions can all be satisfied simultaneously, provides a strong check of the calculation (specifically, these conditions cannot be imposed on the general solution of hard region integral; they are only compatible with the sum of all region integrals, which reproduces the asymptotic expansion of the correlator with four timelike Wilson lines).  

We firstly require the antisymmetry of ${\cal Y}_{(ij;kQ)}^{(l)}$ under $(i,j)$ interchange, 
\begin{align}
  {\cal Y}_{(ji;kQ)}^{(l)}=-{\cal Y}_{(ij;kQ)}^{(l)},
\end{align}
which is a consequence of Bose symmetry. 

Next, we implement the single-pole condition: since $W_{1111}$ consists of connected webs, it does not have subdivergences. Thus, while individual region integrals (which have both infrared and ultraviolet singularities) have multiple poles, their sum must restore the property of the original web, defined with four timelike lines (see Eq.~\eqref{Laurent}) and hence must have a single pole in $\epsilon$. The way this is realised has been analysed in detail in Ref.~\cite{Gardi:2025ule} in two-loop examples. There, all individual region integrals have been computed in full for each web, and then summed up, and contrasted with the expansion of the corresponding web integrals computed directly with timelike lines. 

Next, we consider the boundary condition corresponding to the limit of small squared velocity for all four Wilson lines. In this limit ${\cal Y}_{(ij;kq)}^{(-1)}$ was  computed in Ref.~\cite{Almelid:2015jia,Almelid:2016lrq}. In that calculation, the asymptotic expansion was performed (using the Mellin-Barnes technique) by simultaneously taking all six $\{\alpha_{VU}\}$ (as defined in~(\ref{alphaDef})) variables to zero at the same rate. To this end ${\cal Y}_{(ij;kq)}^{(-1)}$ was first expressed in terms for six variables, two of which remain finite in the limit -- these are $\rho_{ikjq}$ and $\rho_{jkiq}$ (or, equivalently, $\{z,\bar{z}\}$ defined in Eq.~\eqref{zzbVar}) -- as well as four small parameters $\{\alpha_{IK},\alpha_{KQ},\alpha_{JK},\alpha_{JQ}\}$. After the expansion, the latter appear only through logarithms.
To match the result there, we perform a further expansion of our Laurent coefficients ${\cal Y}_{(ij;kQ)}^{(-1)}$ at small $\beta_Q^2$. In this limit the three rescaling invariant variables, $\{r_{ijQ},r_{jkQ},r_{ikQ}\}$, tend to zero with their ratios fixed. According to Eq.~(\ref{newVar}) this lightcone expansion can be realized by sending $x\to 0$. Therefore, we use the following constraint to determine the boundary values:
\begin{align}
\label{gammaexp}
{\cal T}_{x}\left[{\cal Y}_{(ij;kQ)}^{(-1)}(r_{ikQ},r_{jkQ},r_{ijQ};\lambda_i,\lambda_j,\lambda_k)\right]+{\cal O}(x) 
= 
{\cal T}_{\alpha}\left[{\cal Y}_{(IJ;KQ)}^{(-1)}(\rho_{ikjq},\rho_{jkiq};\alpha_{IK},\alpha_{KQ}, \alpha_{JK},\alpha_{JQ})\right] +{\cal O}(\alpha),
\end{align}
where the asymptotic lightcone expansion of the right hand side is in $\alpha$, which is a short-hand notation for the set of four variables,~$\{\alpha_{IK}, \alpha_{KQ}, \alpha_{JK},\alpha_{JQ}\}$. We relate the variables on the two sides of  Eq.~(\ref{gammaexp}) as follows:
\begin{align}
\label{matchingLighcone}
\begin{split}
\frac{z\bar{z}(1-z)(1-\bar{z})}{x(\bar{z}-z)}\sqrt{\lambda_i\lambda_k}=\alpha_{IK},
\qquad
\sqrt{\lambda_k}= \alpha_{KQ},
 \qquad
\frac{z\bar{z}}{x(\bar{z}-z)}\sqrt{\lambda_j\lambda_k}=\alpha_{JK},
      \qquad
      \sqrt{\lambda_j}=\alpha_{JQ}\,,
\end{split}
\end{align}
which can be checked by specifying the definition of the three expansion parameters in eq.~\eqref{virtualities_expansion},
\begin{align}
    \beta_I^2=\frac{4(\beta_I\cdot\beta_Q)^2}{\beta_Q^2}\lambda_i\,,
    \quad 
    \beta_J^2=\frac{4(\beta_J\cdot\beta_Q)^2}{\beta_Q^2}\lambda_j\,,
    \quad
     \beta_K^2=\frac{4(\beta_K\cdot\beta_Q)^2}{\beta_Q^2}\lambda_i\,.
\end{align}
Plugging in Eqs.~\eqref{zzbVar},~\eqref{newVar} and~\eqref{matchingLighcone}, both sides of Eq.~\eqref{gammaexp} depend on 
$\left\{z,\bar{z},x,\sqrt{\lambda_i},
\sqrt{\lambda_j},
\sqrt{\lambda_k}
\right\}$, 
where the four parameters 
$\left\{x,\sqrt{\lambda_i},\sqrt{\lambda_j},\sqrt{\lambda_k}\right\}$ approach zero at the same rate, and, at leading order in the expansion, only appear as logarithms. Therefore, matching the two sides of Eq.~\eqref{gammaexp} provides boundary-value parameters for ${\cal Y}_{(ij;kQ)}^{(-1)}$. 

Next, consider the special configuration of ${\cal Y}_{(ij;kQ)}^{(-1)}$ for the connected webs, where two of the three lightlike velocities, $\beta_u$ and $\beta_v$  become collinear, therefore, the corresponding ratio $r_{u v Q}$ tends to zero, while the remaining two coincide. For example, suppose that the two velocities $\beta_j$ and $\beta_k$ are collinear, then we have~\cite{Liu:2022elt,Duhr:2025cye,Gardi:2025ule},
\begin{align}
\label{dcollVarR}
    \{r_{ijQ}\rightarrow r,\,\,
    r_{jkQ}\rightarrow 0,  \,\,r_{ikQ}\rightarrow r\}\,.
\end{align} 
One can check that by substituting the following parametrization in Eq.~\eqref{newVar},
\begin{align}
\label{CollinearLImit_parametrization}
\begin{split}
    &\left\{z=1-\delta\sqrt{\frac{y (r y-1)}{r (y-1)}},\quad \bar{z}=1-\delta  \sqrt{\frac{y-1}{r y (r y-1)}},\quad x=\delta  \sqrt{\frac{r y-1}{r y\left(y-1\right)}}\right\}, \quad \delta>0,\,\,r<0,\,\, y>1,
    \end{split}
\end{align}
and then sending $\delta$ to zero, the two-particle collinear limit given in Eq.~\eqref{dcollVarR} is reproduced. The three kinematic variables now become $\{\delta,y,r\}$ with $\delta$ measuring how collinear the two velocities are. 

We note that all physical singularities corresponding to $r_{uvQ}\to \{0, 1, \infty\}$ are independent of $y$ at leading order in~$\delta$. In contrast, 
the two symbol letters, $\textcolor{blue}{\Delta_1}$ and $\textcolor{blue}{\Delta_2}$, that appear in the general solution of the differential equation for the connected webs do depend on~$y$ at the leading order in the collinear ($\delta\to 0$) expansion, namely  
\begin{align}
\begin{split}
     \textcolor{blue}{\Delta_1}=\frac{r \left(r y^2-1\right)^2}{(y-1) y (r y-1)}\delta^2 +O(\delta^{3}),
    \qquad
    \textcolor{blue}{\Delta_2}=\frac{r (r (y-2) y+1)^2}{(y-1) y (r y-1)}\delta^2 +O(\delta^{3})\,.
    \end{split}
\end{align}
Other webs contributing to the soft anomalous dimension, only have singularities at $r_{uvQ}\to \{0,1,\infty\}$ and therefore do not involve any dependence on $y$ at leading order in $\delta$. 
Based on strict collinear factorization (Eq.~\eqref{coll}), 
we expect that the collinear limit of the soft anomalous dimension should be smooth and independent of any kinematic variables~\cite{Liu:2022elt,Duhr:2025cye}.  
In particular, the collinear limit of the soft anomalous dimension cannot depend on $y$, and since the other webs are $y$-independent in this limit, this must also be a property of the connected webs ($W_{1111}$) on their own. 
This condition fixes more boundary values, but still leave some boundary terms proportional to Riemann-$\zeta$ values undetermined.
We note in passing that the collinear limit of $W_{1111}$ does depend on $r$ and $\delta$ -- dependence that eventually cancels with the other webs.

The last boundary condition we impose on the solution is the first-entry condition, which requires that only symbol letters corresponding to unitarity cuts in physical channels (Mandelstam invariants) appear as branch points on the principal sheet~\cite{Gaiotto:2011dt,Duhr:2012fh}. 
In the present case, the first-entry condition requires that the symbol must take the form 
\begin{align}
\label{FirstEntry}
     \begin{split}
 {\cal S}\left[ {\cal Y}_{(ij;kQ)}^{(-1)}\right]=r_{ijQ}\otimes\cdots+r_{jkQ}\otimes\cdots+r_{ikQ}\otimes\cdots+\lambda_i\otimes\cdots+\lambda_j\otimes\cdots+\lambda_k\otimes\cdots.
     \end{split}
     \end{align}  
Using the kinematic variables of  Eq.~\eqref{newVar} we therefore require that 
\begin{align}
\label{DiscWeak}
\begin{split}
\text{Disc}_{\omega_n}\left[{\cal Y}_{(ij;kQ)}^{(-1)}\right]=\,\,0,
\qquad 
\forall \omega_n\notin 
    \{x,\bar{z}-z-x,z,\bar{z},1-z,1-\bar{z}\}.
       \end{split}
\end{align}
We find that upon implementing the condition in Eq.~(\ref{DiscWeak}), the result may be expressed as in Eq.~(\ref{FirstEntry}), which is a stronger condition, readily satisfying the Galois symmetry (\ref{GaloisSymm}) in the first entry.

Upon implementing  the first-entry condition, in addition to the conditions discuss thus far, all the boundary values in ${\cal Y}_{(ij;kQ)}^{(-1)}$ are uniquely fixed.

\subsection{Euclidean and physical regimes of the one-mass soft anomalous dimension}

For the quadrupole term ${\cal F}_{1,3}^{(3)}$, there exists a Euclidean regime, 
\begin{align}
    \left\{\beta_i\cdot\beta_j<0,\,\,\, \beta_j\cdot\beta_k<0,\,\,\,\beta_k\cdot\beta_i<0,\,\,\,\beta_i\cdot\beta_Q<0,\,\,\,\beta_j\cdot\beta_Q<0,\,\,\,\beta_k\cdot\beta_Q<0,\,\,\,\beta_Q^2>0\right\},
    \end{align}
    or, in terms of $r_{uvQ}$, 
    \begin{align}
        \left\{r_{ijQ}<0,r_{jkQ}<0,r_{ikQ}<0\right\},
    \end{align}
     where all the three channels of ${\cal F}_{1,3}^{(3)}$ are real.  In contrast, the physical regime, $\{r_{ijQ}>0,r_{jkQ}>0,r_{ikQ}>0\}$, corresponds to all possible cases where $i$, $j$, $k$ and $Q$ are incoming or outgoing particles. Furthermore, ${\cal F}_{1,3}^{(3)}$ takes complex values in the physical regime. 
     
Both the Euclidean and the physical regime correspond to a single-connected domain in the space spanned by real $\{r_{ijQ},r_{jkQ},r_{ikQ}\}$.
However, the function ${\cal F}_{1,3}^{(3)}$ is written in terms of $\{x,z,\bar{z}\}$,  which rationalizes the alphabet. With these variables, the entire Euclidean (physical) regime is divided into seven (five) disconnected domains. This division is related to different signs of the arguments of the two square roots appearing in solving $\{x,z,\bar{z}\}$ from Eq.~\eqref{newVar},
\begin{align}
\label{parametrizing_coll_limit}
\begin{split}
    x=\,\,&\frac{1}{2r_{ikQ}}\left(-\sqrt{\Delta}-\sqrt{\Delta+4r_{ijQ}r_{ikQ}r_{jkQ}}\right),
\\
    z=\,\,&\frac{1}{2r_{ikQ}}\left(r_{ijQ}+r_{ikQ}-r_{jkQ}+\sqrt{\Delta}\right),
\\
    \bar{z}=\,\,&\frac{1}{2r_{ikQ}}\left(r_{ijQ}+r_{ikQ}-r_{jkQ}-\sqrt{\Delta}\right).
    \end{split}
\end{align}
The K\"{a}ll\'{e}n function $\Delta\equiv \Delta(r_{ijQ},r_{ikQ},r_{jkQ})$ is defined by
\begin{align}
    \Delta(x,y,z)\equiv x^2+y^2+z^2-2xy-2yz-2xz.
\end{align}

\subsubsection{Euclidean regime}
We provide the seven disconnected Euclidean domains $R_i^E ,i=1,\cdots,7$, in terms of $\{x,z,\bar{z}\}$:
\begin{itemize}
    \item $\Delta>0,\,\, \Delta+4r_{ijQ}r_{ikQ}r_{jkQ}>0$
\begin{align}
\begin{split}
  \\  R^E_1=&\{0<x<\bar{z}-z,0<z<\bar{z}<1\},
 \\   R^E_2=&\{0<x<\bar{z}-z,1<z<\bar{z}\},
\\   R^E_3=&\{0<x<\bar{z}-z,z<\bar{z}<0\}.
\end{split}
\end{align}
 \item $\Delta>0,\,\, \Delta+4r_{ijQ}r_{ikQ}r_{jkQ}<0$
\begin{align}
\begin{split}
\\   R^E_4=&\{\text{Re}(x)=\frac{1}{2}(\bar{z}-z),\text{Im}(x)>0,0<z<\bar{z}<1\},
\\   R^E_5=&\{\text{Re}(x)=\frac{1}{2}(\bar{z}-z),\text{Im}(x)>0,1<z<\bar{z}\},
\\   R^E_6=&\{\text{Re}(x)=\frac{1}{2}(\bar{z}-z),\text{Im}(x)>0,z<\bar{z}<0\}.
\end{split}
\end{align}
\item $\Delta<0,\,\, \Delta+4r_{ijQ}r_{ikQ}r_{jkQ}<0$
\begin{align}
\begin{split}
\\   R^E_7=&\{\text{Re}(x)=0,\text{Im}(x)>0,\bar{z}=z^*,\text{Im}(z)<0\}\,.
\end{split}
\end{align}
\end{itemize}

\begin{figure}[h]
\begin{subfigure}{0.48\linewidth}
\centering
\begin{tikzpicture}[>=Stealth, node distance=2cm, auto]

  \node (A) {$R_3^E$};
  \node (B) [right=of A] {$R_2^E$};
  \node (C) [above=of $(A)!0.5!(B)$] {$R_1^E$};

\draw[<->] (A) -- node[below] {$T_{i\leftrightarrow k}$} (B);
  \draw[<->] (B) -- node[right] {$T_{j\leftrightarrow k}$} (C);
  \draw[<->] (C) -- node[left] {$T_{i\leftrightarrow j}$} (A);

\end{tikzpicture}
\end{subfigure}
\hfill
\begin{subfigure}{0.48\linewidth}
\centering
\begin{tikzpicture}[>=Stealth, node distance=2cm, auto]

  \node (A) {$R_6^E$};
  \node (B) [right=of A] {$R_5^E$};
  \node (C) [above=of $(A)!0.5!(B)$] {$R_4^E$};

\draw[<->] (A) -- node[below] {$T_{i\leftrightarrow k}$} (B);
  \draw[<->] (B) -- node[right] {$T_{j\leftrightarrow k}$} (C);
  \draw[<->] (C) -- node[left] {$T_{i\leftrightarrow j}$} (A);

\end{tikzpicture}
\end{subfigure}
\caption{Permutation relations between Euclidean domains.}
\label{Fig:PermuEuc}
\end{figure}
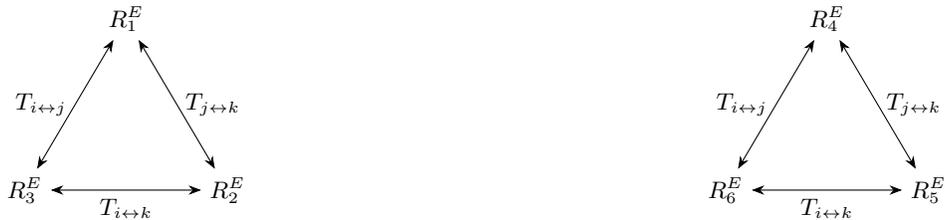

Different Euclidean domains are related by the permutations given in Eq.~\eqref{PermutationRelations}, except for $R_7^E$ where $z$ and $\bar{z}$ are complex conjugate to each other. We depict the action of the permutation on the domains in which the parameters are defined in Fig.~\ref{Fig:PermuEuc}. 
Thus, the permutation relations described in Fig.~\ref{Fig:Permu13} lead to the evaluation of ${\cal F}_{1,3}^{(3)}$ in different Euclidean domains. We checked that ${\cal F}_{1,3}^{(3)}$ is indeed real in all three channels; see Ref.~\cite{MathematicaNotebook}. Note that this is not true for individual GPLs appearing in ${\cal F}_{1,3}^{(3)}$.

\subsubsection{Physical regime}

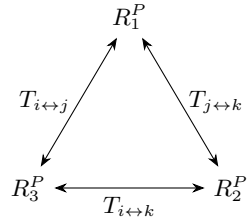
\begin{figure}[h]
\centering
\begin{tikzpicture}[>=Stealth, node distance=2cm, auto]

  \node (A) {$R_3^P$};
  \node (B) [right=of A] {$R_2^P$};
  \node (C) [above=of $(A)!0.5!(B)$] {$R_1^P$};

\draw[<->] (A) -- node[below] {$T_{i\leftrightarrow k}$} (B);
  \draw[<->] (B) -- node[right] {$T_{j\leftrightarrow k}$} (C);
  \draw[<->] (C) -- node[left] {$T_{i\leftrightarrow j}$} (A);

\end{tikzpicture}
\caption{Permutation relations between physical domains.}
\label{Fig:PermuPhy}
\end{figure}

We provide the five disconnected physical domains $R_i^P ,i=1,\cdots,5$, in terms of $\{x,z,\bar{z}\}$:
\begin{itemize}
    \item $\Delta>0,\,\, \Delta+4r_{ijQ}r_{ikQ}r_{jkQ}>0$
\begin{align}
\begin{split}
  \\  R^P_1=&\{0<\bar{z}<z<1,x<\bar{z}-z\},
\\  R^P_2=&\{1<\bar{z}<z,x<\bar{z}-z\},
 \\  R^P_3=&\{\bar{z}<z<0,x<\bar{z}-z\}\,.
 \end{split}
\end{align}
 \item $\Delta<0,\,\, \Delta+4r_{ijQ}r_{ikQ}r_{jkQ}>0$
\begin{align}
\begin{split}
 \\   R^P_4=&\{\text{Re}(x)<0,\text{Im}(x)=-\text{Im}(z),\bar{z}=z^*,\text{Im}(z)>0\}.
 \end{split}
\end{align}
  \item $\Delta<0,\, \, \Delta+4r_{ijQ}r_{ikQ}r_{jkQ}<0$
\begin{align}
\begin{split}
\\   R^P_5=&\{\text{Re}(x)=0,-2\text{Im}(z)<\text{Im}(x)<0,\bar{z}=z^*\}\,.
\end{split}
\end{align}
\end{itemize}
The three real physical domains, $R_1^P$, $R_2^P$ and $R_3^P$, are related by the permutations shown in Fig.~\ref{Fig:PermuPhy}.

\end{document}